\documentstyle[amssymb,11pt]{article}
\begin{document}
\input{epsf}
\title{A new look at relativity transformations}

\author{Ruggiero de Ritis*, Giuseppe Marmo*, Bruno Preziosi**\and
Dipartimento di Scienze Fisiche, Universit\`a di Napoli 
$^{''}$Federico II$^{''}$ \and
*Istituto Nazionale di Fisica Nucleare \and
**Istituto Nazionale di Fisica della Materia}
\maketitle

\begin{abstract}

{\it A free system, considered to be
 a comparison system, allows for the notion of
objective existence and inertial frame. Transformations connecting
inertial frames are shown to be either Lorentz or generalized Galilei.}

\end{abstract}



03.30 - Special relativity.\\

\vskip 3cm

mailing address:\\
Prof. Bruno Preziosi\\
Dipartimento di Scienze Fisiche, Universit\'a di Napoli "Federico II"\\
Mostra d'Oltremare, Pad.20\\
I-80125 Napoli (Italy)\\
tel.n.: (+39) 81-7253416, fax n.: (+39) 81-2394508\\
e-mail: preziosi@na.infn.it\\

\newpage

\section{Introduction}

When modeling physical systems, the carrier space (space of states or
space of events) is usually equipped with some background
mathematical structure (for instance, vector space structure and Euclidean
metric in elementary mechanics, Hilbert space structure and commutation 
relations for quantum mechanics, and so on). With the advent of general 
relativity and the ambition to have a theory of the universe as a whole
it is necessary to minimize the use of
mathematical structures as given a priori. 

Actually the Einstein general approach to physics has this goal:

{\it One of the imperfections of the original relativistic theory of
gravitation was that as a field theory it was not complete: it
introduced the independent postulate that the law of motion of a particle 
is given by an equation of geodesic. A complete theory knows only fields
and not the concepts of particle and motion. For these must not exist 
independently of the field, but are to be treated as part of it} \cite{Ein35}.

 In a footnote  of the same paper, Einstein and Rosen wrote, on the
stress-energy tensor representing the source in the Einstein equations: 

{\it It was clear from the very beginning that this was only a
provisory complexion of the theory in the sense of a phenomenological
interpretation.} 

In any case, no matter what the motivation may be, it seems desirable to
revisit elementary mechanics starting with few assumptions and 
adding  new ones when they are required to resolve ambiguities, if any.
 In this note we would like to consider the minimal
assumptions which may give rise to special relativity and Galilean
relativity and make clear which assumptions will discriminate between
them and what are the compatibilities with the space structure.
It should be remembered that  the  problem of deriving special relativity
without supposing {\it a priori}  the constancy of the light speed  
has been discussed by many authors (some of them are
quoted in ref. \cite{Pre94}).

To better compare different relativity theories, we shall start with
a four-dimensional smooth manifold {\cal M}
as a carrier space for the description
of the evolution of point particles ({\it i.e.}, we start with the space of
events). By using a coordinate system say $(y_{0}, y_{1}, y_{2}, y_{3}
\in {\cal R}^{4})$, we consider the equations of motion in the form
\begin{equation}
\frac{d^{2}y^{\mu}}{ds^{2}}=f^{\mu}(y,\frac{dy}{ds}).
\end{equation}
As a simplifying assumption we require our carrier space to be 
diffeomorphic to ${\cal R}^{4}$.

We are now in a position to define a {\it free system} on our space. The
motion is said to be a free motion if there is a (global) coordinate system
$x^{\mu}=x^{\mu}(y^{\nu})$, such that the equations of motion acquire
the form
\begin{equation}
\frac{d^{2}x^{\mu}}{ds^{2}}=0.
\end{equation}
The parameter $s$ is attached to the particular dynamical system we start with
and has nothing to do with space-time variables.
Any such coordinate system $(x^{\mu})$ will define an affine structure
on the carrier space inducing the one from ${\cal R}^{4}$.
It follows that the notion of straight line, or, more generally, of affine
space on the space of events is {\it frame dependent}. In addition, in 
each frame we must select, first of all,
a coordinate, say $x^{0}$, to be associated with a notion of
{\it evolution}. 

Any solution $x^{\mu}(s)$ of our free equation which represents the 
{\it world-line} of an 
existing object for the given frame must be such that $\frac{dx^{0}}{ds}\ne 0$
along the {\it world-line} (in what follows we shall use $\frac{dx^{0}}{ds}>0$).

After the choice of a single {
\it world-line} has been made, the translation
group can be used to move it and thereby to
 construct a congruence of {\it world-lines}
 which can be thought of 
as solutions of a vector field $E$ on the carrier space. Because the
translation group is a symmetry group for (2), 
these {\it world-lines} are all solutions of (2).

We now  introduce a closed 1-form $\alpha$,  invariant under the translation
group, 
such that $\alpha(E)>0$.  This 1-form $\alpha$ defines a family of 3-planes
for ${\cal R}^{4}$ (a foliation), transversal to the congruence defined by
$E$. As the {\it world-lines} associated with $E$ are solutions of (2), se
say that $(E,\alpha)$ is an inertial frame.
For any such frame, we can define {\it time-like} vectors 
as those $v_{m}\in T_{m}{\cal R}^{4}$ which satisfy $\alpha_{m}(v_{m})\ne 0$.
They will be future pointing if $\alpha(v_{m})>0$. 

We can say that a point $p$ is in the past of $q (p<q)$ if $p$ can be
connected to $q$ by a curve whose tangent vectors are future oriented.
Similarly we can define a point in the future of $q$. These comments 
are meant to stress that $\alpha$ and $E$ allow us to define
most of the standard non metric structures on space-time (if we had given 
a metric structure it would have been possible to define $\alpha$ from 
 $E$ and the metric).

Any {\it time-like world-line}, equipped with a map associating in a monotonic
way a real number to each event on the {\it world-line (clock)} will
be called an {\it observer}. In our approach a clock can be associate to each 
{\it time-like world-line} by considering the pull-back of $\alpha$ to it.

Specifically, this can be done by defining $c_{\alpha}:\gamma({\cal R})
\subset M\rightarrow {\cal R}$ by setting $c_{\alpha}(\gamma(s))=
\int_{s_{0}}^{s}\gamma^{*}(\alpha)$. The induced map $c_{\alpha}:\gamma{\cal R}
\rightarrow {\cal R}$ is clearly monotonic because of $\alpha(E)>0$.

The pair $(E,\alpha)$ defines a family of observers with temporal evolution 
along $E$ and {\it rest frames} associated with $ker~\alpha$. Two inertial 
systems $(E_{1},\alpha_{1})$ and $(E_{2},\alpha_{2})$ are compatible if
$\alpha_{1}(E_{2}))\ne 0$ and $\alpha_{2}(E_{1}))\ne 0$. Two compatible
systems will perceive the {\it world-lines} of each other as representing
some physical entities; for this reason the requirement $\alpha_{a}(E_{b})
\ne 0,
a,b=1,2$ will be called the {\it mutual objective existence} condition.
We shall make the choice $\alpha_{a}(E_{b})>0$.

For any pair of inertial frames, say $(E_{a},\alpha_{a})$,
$(E_{b},\alpha_{b})$, we can define relative clocks by setting
$c_{ba}:{\cal R}\rightarrow {\cal R}$ given by
\[ c_{ba}(s)=c_{b}(\gamma_{a}(s))=\int_{s_{0}}^{s}\gamma_{a}^{*}(\alpha_{b}).\]
Here $s_{0}$ is determined by the intersection point of the two observers
in $(E_{a},\alpha_{a})$ and $(E_{b},\alpha_{b})$ respectively. If the two
observers belong to the same inertial frame, $s_{0}$ is determined by their
intersection with any common rest frame.
This view point makes clear that the notion of
inertial system is related to the dynamical evolution of some chosen 
{\it comparison system} via the selection of a congruence of solutions. 
It should be stressed that with the assumption
that our carrier space is diffeomorphic with ${\cal R}^{4}$, all
{\it free systems} are diffeomorphic to each other. By using the linear
structure induced by an inertial reference  frame, we find that the 
inhomogeneous general linear group ${\cal IGL}(4,{\cal R})$ acts
transitively on the set of solutions of equation (1).

Now we shall look for relativity transformations, {\it i.e.} transformations 
connecting pairs of inertial systems, say $(\alpha_{1},E_{1})$
and $(\alpha_{2},E_{2})$, and require that they form a subgroup of
${\cal IGL}(4,{\cal R})$. 
By selecting one fiducial inertial system  $(\alpha,E)$ we
parametrize all others connected to it by the elements of the relativity group
we are searching for. 

The restrictions on the accepted 
relativity transformations
means that a {\it particle} at rest in one inertial frame
cannot be perceived by another one as a particle {\it existing} all
over a real line only at a given instant of time, \underline {without past}
 and \underline {without future}.

From the point of view of the transformation group ({\it i.e.} transformations 
connecting physically equivalent systems) we have to exclude the possibility
of exchanging {\it time-} and {\it space-}
 axes. For instance, we should exclude from
admissible {\it relativity transformations} those closing on a
subgroup of ${\cal IGL}(4,{\cal R})$ which contains
rotations in the {\it  time-space} planes.

We shall now describe how to construct these {\it relativity (sub)-groups}.
The main idea of our procedure consists of looking for the  transformation
groups which transform an inertial system into another one 
compatible with the {\it mutual objective existence} condition. 
This is  translated
in the requirement that a {\it transformed $\alpha$}, say $\tilde{\alpha}$
 should not contain $E$
in the kernel, {\it i.e.} $\tilde{\alpha}(E)\ne 0$ and $\alpha(\tilde{E)}\ne 0$.

Before carrying on this program we shall consider some preliminary aspects
(sections 2-4).

\section {On the conditions for a system to be free}

Given a second order differential equation on space-time, say
\begin{equation}
\frac{d^{2}y^{\mu}}{ds^{2}}=f^{\mu}\left(y,\frac{dy}{ds}\right),
\end{equation}
we ask under which conditions we can find a new coordinate system
in which the equation becomes 
\begin{equation}
\frac{d^{2}x^{\mu}}{ds^{2}}=0.
\end{equation}
Clearly, by performing a change of coordinates $x^{\mu}=x^{\mu}(y)$, we find:

\[
0=\frac{d}{ds}\left(\frac{dx^{\mu}}{dy^{\nu}}\frac{dy^{\nu}}{ds}\right)=
\frac{d^{2}x^{\mu}}{dy^{\nu}dy^{\rho}}\frac{dy^{\rho}}{ds}\frac{dy^{\nu}}{ds}
+\frac{dx^{\mu}}{dy^{\nu}}\frac{d^{2}y^{\nu}}{ds^{2}},
\]
{\it i.e.}

\[
\frac{d^{2}y^{\nu}}{ds^{2}}=
\left(\left(\frac{dx}{dy}\right)^{-1}\right)^{\nu}_{\mu}
\frac{d^{2}x^{\mu}}{dy^{\alpha}dy^{\beta}}\frac{dy^{\alpha}}{ds}
\frac{dy^{\beta}}{ds}=\Gamma^{\nu}_{\alpha\beta}\frac{dy^{\alpha}}{ds}
\frac{dy^{\beta}}{ds}.
\]
 
If we compute the curvature associated with this connection, we find

\[
\cal{R}^{\lambda}_{\nu\mu\rho}=\frac{\partial \Gamma^{\lambda}_{\mu\nu}}
{\partial y^{\rho}}-\frac{\partial \Gamma^{\lambda}_{\mu\rho}}
{\partial y^{\nu}}+\Gamma^{\eta}_{\mu\nu}\Gamma^{\lambda}_{\rho\nu}
-\Gamma^{\eta}_{\mu\rho}\Gamma^{\lambda}_{\nu\eta} = 0;
\]
as a matter of fact this condition turns out to be also sufficient
to go from (3) to (4).

As in general we are not requiring the force $f^{\mu}$ in (11) to be quadratic
in the velocities, this sufficiency condition should be stated in the
framework of generalized connections associated with any second order 
vector field \cite{Cra96} \cite{Mor90}:

\begin{equation}
R_{ij}^{m}= -\frac{1}{2}\left[\frac{\partial^{2}f^{m}}{\partial q^{i}
\partial u^{j}}-\frac{\partial^{2}f^{m}}{\partial u^{i}
\partial q^{j}}+\frac{1}{2}\left(\frac{\partial^{2}f^{m}}{\partial u^{j}
\partial u^{l}}\frac{\partial f^{l}}{\partial u^{i}}-
\frac{\partial^{2}f^{m}}{\partial u^{i}
\partial u^{l}}\frac{\partial f^{l}}{\partial u^{j}}\right)\right]=0.
\end{equation}
In the particular case our starting system is in the form of a spray

\[
f^{i}=-\Gamma^{i}_{km}u^{k}u^{m},
\]
we have

\[
\Gamma^{k}_{ij}=-\frac{1}{2}\frac{\partial^{2}f^{k}}{\partial u^{i}
\partial u^{j}}, ~~~~~   R^{m}_{ij}=\tilde{R}^{m}_{kij}u^{k} .
\]
Thus, relation (5) is necessary and sufficient condition for a second
order differential equation to represent a free system on ${\cal R}^{4}$.

In coming section we shall describe a constructive procedure to find
these special reference frames.

\section{{\it Natural coordinates} for second order equations:
Searching for inertial frames}
 Starting with a second order equation on $M$, say

\[
\frac{d^{2}y^{\mu}}{ds^{2}}=f^{\mu}(y,\frac{dy}{ds}),
\]
we can define a {\it natural coordinate system}
 for any neighbourhood $U\ni m$
in the following way.

We consider a fiducial point $m$ and the tangent space $T_{m}M$.
With any vector $v_{m}\in T_{m}M$ we associate a point in $M$
by looking for the (unique) solution of (1) originating at $m$ with
initial velocity $v_{m}$.

The flow $\phi:{\cal R}\times TM\rightarrow TM$, associated with (1),
defines a map by restriction

\[
\begin{array}{l}
\phi:{\cal R}\times \{m\}\times T_{m}M\rightarrow U\subset M,\\
\phi(s,m,v_{m})=m_{v}(s),
\end{array}
\]
where $m_{v}(s)$ is the point in $U$ reached after a {\it time} $s$ via the
solution of (1) with initial conditions $\{m,v_{m}\}$. In particular
the application $\phi(s=1,m,v)=m_{v}$ may be considered as a
generalization to a generic second order equation
of the standard exponential map for geodesic equations.

For simplicity, we assume that (1) defines a complete vector field on
$TM$. In this hypothesis the particular map we have constructed is defined
for any $v_{m}\in T_{m}M$ injectively
(we recall that $M$ is
diffeomorphic to ${\cal R}^{4}$ by assumption and our considerations apply
to a neighbourhood $U$ of $m$). In particular we may define an
addition rule on $U$ by setting $m_{v_{1}}+m_{v_{2}}:=m_{v_{1}+v_{2}}$.

The vector space structure induced on $U$ for each, complete, second order
vector field on $M$, depends on $m$. When these vector space structures on
$U$ are linearly related, {\it i.e.} transition functions are linear maps, our
starting equation (1) reduces to a {\it free particle} equation.
When moving from a point $m$ to a point $m'$ we go from the vector space 
structure associated with $T_{m}M$ to the one associated to $T_{m'}M$,
this can be done {\it transporting} one vector space onto the other
along the {\it solution curve} connecting $m'$ to $m$.

Depending on $m$ and $m'$ this connection map may fail to be {\it linear}
up to some order depending on the {\it extension} of the neighbourhood, the set 
of points we may reach from $m_{0}$ by using {\it solution curves} while
keeping the {\it linearity} violated to no more than some preassigned
power $k$ in the parameters will be a $k$-order local inertial frame.

When equation (5) is satisfied, this procedure defines a global linear
inertial frame. 

\section{Transforming inertial frames}

We start with an inertial frame, {\it i.e.} a
 reference inertial frame described by $(E,\alpha)$, 
giving rise to a global coordinate system $(x^{0},x^{1},x^{2},x^{3})$.
We notice that a particular form for $\alpha$ would be $\alpha=dx^{0}$.

It is worth stressing that a different choice of the congruence ({\it i.e.} a
different choice of $E$) in general may give 
rise to a different class of inertial systems.

Having found one coordinate system in which the equation of motions have
the form (2), how many of them exist?

It is clear we have to look for all coordinate systems $\xi^{\mu}=
\xi^{\mu}(x,v)$ such that

\[
\frac{d^{2}\xi^{\mu}}{ds^{2}}=0.
\]

As

\[
\frac{d\xi^{\mu}}{ds}=\frac{\partial\xi^{\mu}}{\partial 
x^{\nu}}\frac{dx^{\nu}}{ds}
+\frac{\partial\xi^{\mu}}{\partial v^{\nu}}\frac{dv^{\nu}}{ds}, 
\]
by using

\[
\frac{dx^{\mu}}{ds}=v^{\mu},~~~~\frac{dv^{\mu}}{ds} = 0 
\]
we find

\[
\frac{\partial^{2}\xi^{\mu}}{\partial x^{\nu}\partial x^{\rho}} 
v^{\rho} v^{\nu} = 0,  
\]
{\it i.e.}, $\xi^{\mu}$ must be linear in $x^{\nu}$
(it can be however any function of constants of the motion
for system (2)). We 
consider therefore

\[
\xi^{\mu}=A^{\mu}_{\nu}(v)x^{\nu}+a^{\mu}(v),~~~~~\frac{d\xi^{\mu}}{ds}=
A^{\mu}_{\nu}(v)\frac{dx^{\nu}}{ds}=w^{\mu}.
\]

In terms of the initial conditions, we have
\begin{eqnarray}
&\xi^{\mu}(s)=A^{\mu}_{\nu}(v(0))x^{\nu}(0)+A^{\mu}_{\nu}(v(0))v^{\nu}(0)s 
+a^{\mu}(v(0))=\nonumber\\
&=A^{\mu}_{\nu}(v)x^{\nu}+ A^{\mu}_{\nu}(v)v^{\nu}s + a^{\mu}(v) = 
\xi^{\mu}(0)+w^{\mu}(v)s,\nonumber
\end{eqnarray}
where $v$ stays for $v(s)$. 

{\it \underline{Remark}

 As constants of the motion, say \cal{C}, for our comparison dynamics
satisfy $\frac{d}{ds}\cal{C}=0$, it is clear that $\xi^{\mu}$ can be any
function of them in addition to the explicit dependence on ($x^{\mu}$).
Here, we only consider the dependence on constants of the motion ($v^{\mu}$)
and do not consider a possible dependence on other constants of
motion. This
is a simplifying assumption useful for computations because
 $(x^{\mu},v^{\mu})$ parametrize the
position-velocity phase space.}

Because our approach to inertial frames is a dynamical one, there is
 no reason to restrict
our transformations to be {\it point transformations} 
({\it i.e.} to tangent bundle
automorphisms, to use the language from differential geometry). Therefore,
{\it new} velocities need not be linear functions of the {\it old}
velocities. The congruence of curves corresponding to

\[
x^{\mu}(s)=v^{\mu}s + x^{\mu}(0),~~~~~~v^{0}>0
\]
will be given by

\[
\xi^{\mu}(s)=w^{\mu}s+\xi^{\mu}(0),~~~~~~w^{0}>0,
\]
where ($w^{\mu}$) can be any smooth function of $(v^{\mu})$.

\section{Relativity transformations}

\subsection{A preliminary lemma} 

At this point we look for linear transformations on a given pair 
$(E,\alpha)$ with the requirement that any transformed one 
still satisfies $\tilde{\alpha}(E)>0$, $\alpha(\tilde{E})>0$, {\it i.e.}
our transformations 
generate new inertial frames satisfying the {\it mutual objective existence}. 
From here we shall be able to construct the relativity groups which
are compatible with our
requirement.

As usual, we can dispose of the translation part and concentrate our analysis
on the linear homogeneous part.

By using a passive point of view, we can consider our transformations
from ${\cal R}^{4}$ to ${\cal R}^{4}$, preserving the origin. 

We shall therefore consider linear transformations $x^{\mu}=A^{\mu}_{\nu}
\tilde{x}^{\nu}$,
with the additional requirement $\frac{dx^{0}}{d\tilde{x}^{0}}>0$, to 
implement $\tilde{\alpha}(E)>0$, $\alpha(\tilde{E})>0$.  Here we think of the
choice $E=\frac{\partial}{\partial x^{0}}, \alpha=dx^{0};
\tilde{E}=\frac{\partial}{\partial \tilde{x}^{0}}, 
\tilde{\alpha}=d\tilde{x}^{0}$.

We have a preliminary lemma:

{\it Any invertible 
linear transformation,} ${\cal R}^{4}\rightarrow {\cal R}^{4}$,
{\it which, along with its inverse, 
preserves the time-like character can be decomposed into the 
product of a linear transformation in the} $(0,1)${\it -plane and two
space-like transformations}.

We denote by $A$ our generic transformation and apply it to 
a standard vector:

\[
A
\left(\begin{array}{c}
1\\0\\0\\0 \end{array}\right)=
\left(\begin{array}{c}
a_{0}\\a_{1}\\a_{2}\\a_{3} \end{array}\right)
\]
where $a_{0}> 0$ by assumption.

Now, by using a space-transformation $R$, we can transform
\[\left(\begin{array}{c}
a_{1}\\a_{2}\\a_{3} \end{array}\right)~~~~into~~~~
\left(\begin{array}{c}
a\\0\\0 \end{array}\right).\]
Therefore
\[
RA\left(\begin{array}{c}
1\\0\\0\\0 \end{array}\right)=
\left(\begin{array}{c}
a_{0}\\a\\0\\0 \end{array}\right).
\]
By using a linear transformation $L$ in the $(0,1)$-plane,
acting as the identity in the remaining components and preserving
the {\it time-like} character, we find

\[
LRA\left(\begin{array}{c}
1\\0\\0\\0 \end{array}\right)=
\left(\begin{array}{c}
1\\0\\0\\0 \end{array}\right).
\]

By using the arbitrariness of the starting time-like vector, 
we find that $S\equiv LRA$ is a space-transformation and
we get the decomposition

\[
A = R^{-1}L^{-1}S .
\]

This decomposition lemma allows us to deal first with transformations in the
$(0,1)$ plane to find out which ones are compatible with the requirement on
the {\it objective existence} condition and then to compose them. 
This analysis will be done in the following sections.

\subsection{Infinitesimal transformation and associated quadratic forms}

Having reduced our problem to a two-dimensional one, it is  easy to
visualize the situation. A vector $E$ is given and we should consider 
all those linear transformations in ${\cal R}^{2}$ which never take a
vector transversal to $E$ (defined by $ker~\alpha$) into one parallel to $E$.
If we think of a one-parameter group of transformations connecting two
allowed frames,  for any infinitesimal generator ${\cal A}$, there
will never be a value of the parameter $\sigma$ for which 
$e^{\sigma{\cal A}}~ ker~\alpha$
is parallel to $E$. 

At this point it is convenient to decompose any element of $GL(2,{\cal R})$ 
into the product of an element in $SL(2,{\cal R})$ and a dilation. Because
dilations are in the center of $GL(2,{\cal R})$ they can be dealt with 
separately. Thus, we may restrict our analysis to $SL(2,{\cal R})$.

For this analysis it is very convenient to notice that $SL(2,{\cal R})$ is 
the same as $Sp(2,{\cal R})$,{\it i.e.}  the group of canonical transformations
in a two-dimensional phase-space. From this point of view,
 a matrix ${\cal A}$ is associated
with some Hamiltonian function whose level sets contain the orbits of the
one-parameter group associated with ${\cal A}$ \cite{Hir74}.

It is now clear that the family of inertial frames we 
obtain from a given one with 
the action of the one-parameter group $e^{\sigma{\cal A}}$, is 
associated with a quadratic form, the Hamiltonian function 
{\it generating} ${\cal A}$, up to a numerical factor. Thus we are led to
analyse quadratic forms on ${\cal R}^{2}$ in connection with the 
{\it placement} of $(E,ker~\alpha)$ in space-time, keeping in mind
that their level sets contain the orbits of the {\it relativity transformation
group} we are searching for. 

The infinitesimal transformation ${\cal A}$ associated 
with a transformation matrix
\begin{equation}
A=a\left(\begin{array}{cc}1-\tilde{\epsilon} &\tilde{\alpha}_{01}\\ 
\tilde{\alpha}_{10}&
1+\tilde{\epsilon}\end{array}\right),
\end{equation}
which tends to the unit of the group when its parameters tend to zero, has
the form

\[\left(\frac{dA}{d\sigma}\right)_{\sigma=0}=
{\cal A}=\left(\begin{array}{cc}-\epsilon &\alpha_{01}\\ \alpha_{10}&
\epsilon\end{array}\right).
\]
The evolution of the vector  $\left(\begin{array}{c}x^{0}\\x\end{array}
\right)$ is described by
\begin{equation}
\frac{d}{d\sigma}\left(\begin{array}{c}x^{0}\\x\end{array}
\right)={\cal A}\left(\begin{array}{c}x^{0}\\x\end{array}
\right)=
\left(\begin{array}{c}\alpha_{01}x-\epsilon x^{0}\\
\alpha_{10}x^{0}+\epsilon x\end{array}\right)=
\left(\begin{array}{c}-\frac{\partial H}{\partial x}\\
\frac{\partial H}{\partial x^{0}}\end{array}\right),
\end{equation}
where
\[
2H=\alpha_{10}(x^{0})^{2}+2\epsilon x^{0}x-\alpha_{01}x^{2}\]
If ${\cal A}$ does not admit real eigenvalues, it  will  define a rotation-like
transformation ({\it i.e.} level sets of $H$ are ellipses) and will violate
the {\it mutual objective existence} (in what follows {\it m.o.e.}) 
condition. We
assume therefore that ${\cal A}$ has real eigenvalues. We get:
\[
-\epsilon x^{0}+\alpha_{01}x=\lambda x^{0},~~~~
\alpha_{10}x^{0}+\epsilon x=\lambda x,~~~~
\lambda^{2}\equiv\epsilon^{2}+\alpha_{01}\alpha_{10}.\]
Eigen-directions are defined by
\[\alpha_{10}x^{0}+(\epsilon\pm \lambda)x=0\]
and, in terms of them, we may write
\begin{equation}
2H=\alpha_{10}
\left(x^{0}+\frac{x}
{c_{-}}\right)
\left(x^{0}-\frac{x}{c_{+}}\right),~~~~where~~~
\frac{1}{c_{\mp}}=\frac{|\lambda|\pm\epsilon}
{\alpha_{10}}.
\end{equation}
The solution of eq. (7)  corresponding to the initial conditions 
$x^{0}=x^{0}_{0},~x=x_{0}$ is  given by
\begin{equation}
\left(\begin{array}{c}x^{0}\\x\end{array}\right)=
\cosh \sigma\lambda\left(\begin{array}{cc}
1-\epsilon\frac{\tanh \sigma\lambda}
{\lambda}
&\alpha_{01}\frac{\tanh \sigma\lambda}{\lambda}\\
\alpha_{10}\frac{\tanh \sigma\lambda}
{\lambda}&
1+\epsilon\frac{\tanh \sigma\lambda}
{\lambda}\end{array}\right)
\left(\begin{array}{c}x^{0}_{0}\\x_{0}\end{array}\right).
\end{equation}

Moreover, the ratio $\frac{x}{x^{0}}$, when
$\sigma\rightarrow \pm\infty$, tends  to $(-c_{-},c_{+})$, independently of the
value of the initial conditions. As a consequence,
these quantities are characteristic invariants
of the transformation. The quantities 
\[\frac{\alpha_{01}}{\alpha_{10}}=\frac{1}{c^{2}}\equiv \frac{1}{c_{-}c_{+}}
~~~~and~~~~\frac{\epsilon}{\alpha_{10}}=\frac{1}{c_{1}}=\frac{1}{2}
\left(\frac{1}{c_{-}}-\frac{1}{c_{+}}\right)\]
are also invariant. Finally, with the substitution
$\tanh\sigma\lambda=\frac{\tilde{\alpha}_{10}}{2}\left(\frac{1}{c_{-}}+ 
\frac{1}{c_{+}}\right)$,
the transformation matrix in (9) takes the form
\begin{equation}
A=\frac{1}{\sqrt{1-\tilde{\alpha}^{2}_{10}
\left(\frac{1}{c_{1}^{2}}+\frac{1}{c^{2}}
\right)}}\left(\begin{array}{cc}1-\frac{\tilde{\alpha}_{10}}{c_{1}}&
\frac{\tilde{\alpha}_{10}}{c^{2}}\\
\tilde{\alpha}_{10}&1+\frac{\tilde{\alpha}_{10}}{c_{1}}\end{array}\right).
\end{equation}
Here $\tilde{\alpha}_{10}$ plays the role of parameter of the transformation; 
no ambiguity arises if we continue to indicate it with $\alpha_{10}$.

The level set corresponding to $H={\overline H}=0$ determines the asymptotes
of the branches of the hyperbolas defined by $H={\overline H}>0$  and
$H={\overline H}<0$. The asymptotes
 coincide with the {\it eigen-directions} of $A$. 

It is now clear that $A$, associated to $H$, will define an acceptable
relativity transformation only if $ker~\alpha$ intersects the branches
corresponding to ${\overline H}<0$ and $E$ the branches corresponding to
${\overline H}\geq 0$ (or viceversa, and then we redefine the group parameter
so that $ker~\alpha$ always intersects the negative branch). This implies
that $A$ (see eq. (6)) cannot transform a vector whose second component 
is zero in a vector whose first component is zero.

 A priori, for
a given pair $(E,\alpha)$, we shall find several quadratic functions $H$
satisfying previous requirements, therefore we may think that a combination
of them would be also acceptable. Which combinations may be admissible will
be discussed in next subsection. 

\subsection{Generic Hamiltonians and compatible transformations}

We analyse the trajectories associated with a generic Hamiltonian function 
(8) $(\alpha_{10}\geq 0)$
in the $(x^{0},x)$ plane, with reference to Fig. 1.

\vskip 5mm

{\bf  {\it a)}} $\alpha_{10}\alpha_{01}<-\epsilon^{2}$;  

no compatibility with the {\it m.o.e.}
condition; this is the rotation-like case.

\vskip 5mm
{\bf {\it b)}} $\alpha_{10}\alpha_{01}=-\epsilon^{2}$; 

the Hamiltonian reduces to $\alpha_{10}(x^{0}+\frac{x}{c_{1}})^{2}$:
no compatibility with the {\it m.o.e.}
condition;

\vskip 5mm
{\bf {\it c)}} $-\epsilon^{2}<\alpha_{10}\alpha_{01}<0$;

both eigen-directions are contained between $E$ and $ker~\alpha$;
no compatibility with the {\it m.o.e.}
condition;

\vskip 5mm
{\bf {\it d)}} $\alpha_{10}\alpha_{01}>0$;

this condition  selects candidates to be acceptable Lorentz 
transformations.

{\it i)} Two transformations $A$ and $A'$ of the form (10)
belong to the same transformation 
group only if they have identical
invariants; if this is not the case, the {\it m.o.e.} existence is violated
by the transformation  we get composing  some powers of them;
in fact, if, {\it e.g.}, $\frac{1}{c'_{+}}>\frac{1}{c_{+}}$, a {\it world-line},
admissible for $A'$, may be rotated by some power of $A$ into $ker~\alpha$
(for the same reason Lorentz-type transformations will not be
 compatible with Carroll and/or 
Galilei transformations).   

{\it ii)}
In addition, the product is commutative if and only if 
$\alpha_{01}\alpha'_{10}=\alpha'_{01}\alpha_{10}$
and 
$\epsilon'\alpha_{01}=\epsilon\alpha'_{01}$,
that is to say if 
and only if the
{\it m.o.e.} condition is satisfied.

{\it iii)} The addition rule for the unique parameter is:  
\[\alpha^{''}_{10}=
\frac{\alpha_{10}+\alpha'_{10}}{1+\alpha_{10}\alpha'_{10}
\left(\frac{1}{c_{1}^{2}}+\frac{1}{c^{2}}\right)}\].
 
\vskip 5mm

{\bf {\it e)}} $\alpha_{10}=0,~ 2H=x(2\epsilon x^{0}-\alpha_{01}x)$;

in this {\it Carroll} case \cite{Lev76}

\[A=\frac{1}{\sqrt{1-\epsilon^{2}}}\left(\begin{array}{cc}
 1-\epsilon & \alpha_{01}\\ 0 & 1+\epsilon \end{array} \right);
\]

{\it i)} $E$ is an invariant asymptote; no transformation can move $E$ 
into $ker~\alpha$; 

{\it ii)} {\it equi-locality} is absolute ($dx_{0}=0 \rightarrow dx=0$).

\vskip 5mm

{\bf {\it f)}} $\alpha_{01}=0,~ 2H=x^{0}(\alpha_{10}x^{0}+2\epsilon x)$;

in this Galileian case

\[A=\frac{1}{\sqrt{1-\epsilon^{2}}}\left(\begin{array}{cc}
 1-\epsilon & 0\\ \alpha_{10} & 1+\epsilon \end{array} \right);
\]

{\it i)} $ker~\alpha$ is an invariant asymptote,

{\it ii)} the product of two transformations of the same type,(with different values
for $\epsilon$ and $\alpha_{10}$) satisfies also
the {\it m.o.e.} condition;

{it iii)} {\it simultaneity} is absolute ($dx^{0}_{0}=0 \rightarrow dx^{0}=0$).

\vskip 5mm

{\bf {\it g)}} $\alpha_{01}=\alpha_{10}=0;~ H=\epsilon x^{0} x$;

this Aristotelian transformation preserves both {\it equi-locality}
 and {\it simultaneity} and
is compatible with Lorentz, Galilei and Carroll.

\subsection{Requiring the identity of relative clocks}

Inertial frames connected by a relativity transformation should be
considered to be equivalent; therefore we make a further requirement.
We impose that inertial frames
connected by an allowed relativity transformation should have identical 
relative clocks, {\it i.e.} we require that the two maps $c_{ab}$ and $c_{ba}$
coincide for any two inertial frames we obtain starting with $(E,\alpha)$. 
This requirement will impose $\epsilon=0$. Therefore our analysis is now
greatly simplified.

The argument for $\epsilon=0$ is simple; eq. (9) implies that previous
condition is satisfied if
\begin{equation}
\frac{\partial x^{'0}}{\partial x^{0}}=A_{00}=
\frac{\partial x^{0}}{\partial x^{'0}}=
(A^{-1})_{00},
\end{equation}
that is to say if $c_{1}\rightarrow \infty,~ \epsilon=0$.
This equation, which is independent of the restriction to a bidimensional 
space-time, implies that dilation along the time-axis should be excluded from 
our relativityu transformations, {\it i.e.} our infinitesimal transformations 
should not contain $x^{0}\frac{\partial}{\partial x^{0}}$.

Going back to a bi-dimensional space-time, we find that
the Hamiltonian (8) assumes the form
\begin{equation}
2H=\alpha_{10}\left((x^{0})^{2}-\frac{x^{2}}{c^{2}}\right)=
\alpha_{10}\left(x^{0}+\frac{x}{c}\right)\left(x^{0}-\frac{x}{c}\right),
\end{equation}
while  $A$  reduces to
\begin{equation}
A=\frac{1}{\sqrt{1-\frac{\alpha^{2}_{10}}{c^{2}}}}\left(\begin{array}{cc}
1&\frac{\alpha_{10}}{c^{2}}\\ \alpha_{10}&1\end{array}\right).
\end{equation}

We conclude that our (1,1) {\it space-time} imay be equipped with an invariant 
quadratic form given by $dx^{0}\otimes dx^{0}-\frac{1}{c^{2}}dx\otimes dx$
in the (generic) Lorentz case and by $dx^{0}\otimes dx^{0}$ in the Galileian
 case. 

The physical interpretation of these transformations and the related problems,
like the clock synchronization one, have been extensively discussed in the
literature \cite{Rei28},  \cite{Gru73}, \cite{Hav87}.

We conclude this section by remembering that if space and time coordinates of
the universe are deformed in such a manner that all the space-time 
coincidences are conserved, then the universe remains unchanged \cite{Sch22}.
Notice that this requirement is satisfied already at the level in which only
the {\it m.o.e.} is supposed.

\section{Back to four-dimensional space-time}   

We have seen that, by using the decomposition lemma in section 
5, we have been able to select transformations compatible with
our requirement of equality of relative clocks and {\it 
mutualobjective existence} in the $(0,1)$-space time.

Here we would like to find out which are the implications in four dimensions,
if we compose it with space transformations.

We first consider the $(1,1)$-Galilei group. It is clear that because 
our transformations preserve a space-slicing ({\it i.e.} an absolute notion
of simultaneity), any linear transformation along the space part will be
an acceptable relativity transformation. We find for the homogeneous 
{\it generalized} Galilei group the semidirect product 
\begin{equation}
G_{0}~:=~{\cal V}\times_{\rho}GL(3,{\cal R}),   \nonumber
\end{equation}
where ${\cal V}$ stays for the three dimensional space of velocities.

By defining the action on a vector space-time $(b,\vec{x})$ we find
\begin{equation}
(\vec{v},A)[(b,\vec{x})]=[(b,A\vec{x}+b\vec{v})]
\end{equation}
along with the composition rule
\begin{equation}
(\vec{v}_{1},A_{1})\cdot(\vec{v}_{2},A_{2})=(A_{1}\vec{v}_{2}+\vec{v}_{1},
A_{1}\cdot A_{2}).\end{equation}

The rotation group of the standard Galilei group is being replaced by
the General Linear group in three dimensions, i. e. the {\it m.o.e.} condition
does not require any preferred notion of distance along the space part
of space-time. A symmetric $(0,2)-$tensor of the type $dx^{0}\otimes dx^{0}$
is invariant under the action of $G_{0}$.
Now we consider the other compatible group, {\it i.e.} the {\it Lorentz type}
transformations. Here, to carry on computations,
 we find more convenient to use infinitesimal
transformations in terms of vector fields.

For each plane involving a time coordinate and a space coordinate, our 
$(1,1)$ analysis provides us with the following vector fields:
\begin{eqnarray}
&B_{1}=\alpha^{-1}_{01}\alpha_{10}x_{1}\frac{\partial}{\partial x_{0}}+
x_{0}\frac{\partial}{\partial x_{1}}, \nonumber\\
&B_{2}=\alpha^{-1}_{02}\alpha_{20}x_{2}\frac{\partial}{\partial x_{0}}+
x_{0}\frac{\partial}{\partial x_{2}}, \nonumber\\
&B_{3}=\alpha^{-1}_{03}\alpha_{30}x_{3}\frac{\partial}{\partial x_{0}}+
x_{0}\frac{\partial}{\partial x_{3}}.\nonumber
\end{eqnarray}

It is convenient to redefine coordinates by setting 
$y_{1}=\sqrt{\alpha_{01}\alpha_{10}}x_{1},
y_{2}=\sqrt{\alpha_{02}\alpha_{20}}x_{2},
y_{1}=\sqrt{\alpha_{03}\alpha_{30}}x_{3},
y_{0}=x_{0}$;
we find then
\begin{eqnarray}
&\tilde{B}_{1}=y_{1}\frac{\partial}{\partial y_{0}}+
y_{0}\frac{\partial}{\partial y_{1}}, \nonumber\\
&\tilde{B}_{2}=y_{2}\frac{\partial}{\partial y_{0}}+
y_{0}\frac{\partial}{\partial y_{2}}, \nonumber\\
&\tilde{B}_{3}=y_{3}\frac{\partial}{\partial y_{0}}+
y_{0}\frac{\partial}{\partial y_{3}}.\nonumber
\end{eqnarray}
Now we look for space transformations whose commutator with $\tilde{B}$'s
does not violate the {\it m.o.e.} condition.

We find
\begin{eqnarray}
& \left[ A^{i}_{j}y^{j}\frac{\partial}{\partial y^{i}},\tilde{B}_{k}\right]=
A^{k}_{j}y^{j}\frac{\partial}{\partial y_{0}}
-y_{0}A^{j}_{k}\frac{\partial}{\partial y^{j}}=\nonumber\\
&=A^{k}_{j}\left(y^{j}\frac{\partial}{\partial y_{0}}+
y_{0}\frac{\partial}{\partial y^{j}}\right)
-(A^{k}_{j}+A^{j}_{k})y_{0}\frac{\partial}{\partial y^{j}},\nonumber
\end{eqnarray}
{\it i.e.} we find a combination of {\it boosts} and Carrol transformations.
To preserve the {\it m.o.e.} condition we have to require
$A^{k}_{j}=-A^{j}_{k}$; in conclusion, the most general space transformations
compatible with {\it boosts} to preserve the objective conditions are
just rotations.

We find that our allowed relativity group is the Lorentz group. Therefore
our space time gets equipped with a generalized Minkowskian metric.
The associated symmetric $(0,2)-$tensor has the form $dx^{0}\otimes dx^{0}-
\beta_{1}^{2} dx^{1}\otimes dx^{1}-\beta_{2}^{2} dx^{2}\otimes dx^{2}-
\beta_{3}^{2} dx^{3}\otimes dx^{3}$.

Some further intermediate situations are possible. They correspond to the 
use of Galilei-type trasnformations in some {\it one time-ome space}
planes and Lorentz type in the remaining ones. These situatiomns are obtained
when some of the coefficients of the rprevious quadratic form are being set 
equal to zero, say $\beta_{1}=0, \beta_{2}$ and $\beta_{3}$ being different
from zero, or $\beta_{1}=\beta_{2}=0$, $\beta_{3}$ being different from zero. 
\section {Conclusions}

We have found that the notion of {\it mutual objective existence} along
with the identity of relative clocks is enough 
to select only the Galilei and Lorentz transformations in one space and
one time setting. These transformations preserve a quadratic form which is
degenerate in the Galileian case and not degenerate in the Lorentz case.

When going to $\cal{R}^{4}$, we find that the {\it mutual  
objective existence} condition
determines a Minkowski-type metric in the case of the Lorentz group, while 
the Galilei-type group does not impose restrictions on the space structure.

Some intermediate situations are also possible. It is possible to have 
Lorentz-type behaviour in some directions and Galilei-type behaviour in
the complementary directions. These mixed situations cannot be ruled out
only on the basis of the {\it mutual objective existence} condition and
some additional physical insight is needed.

We hope we have made clear how a {\it general relativity ideology} may be 
useful in dealing also with {\it special relativity}, where only an affine
structure for space-time  is needed.

\section{Acknowledgements}
The authors are grateful to W. M. Tulczyjew for useful discussions, to
 E.J. Saletan for reading the manuscript
and to the partecipants of the {\it Fryday seminars}
where these results were presented in several occasions and helped us to
clarify several subtle points. Thanks are also due to Alessandro Preziosi
(9 years old, grandchild of one of the authors) for his computer drawing.

\begin{figure}
\epsffile{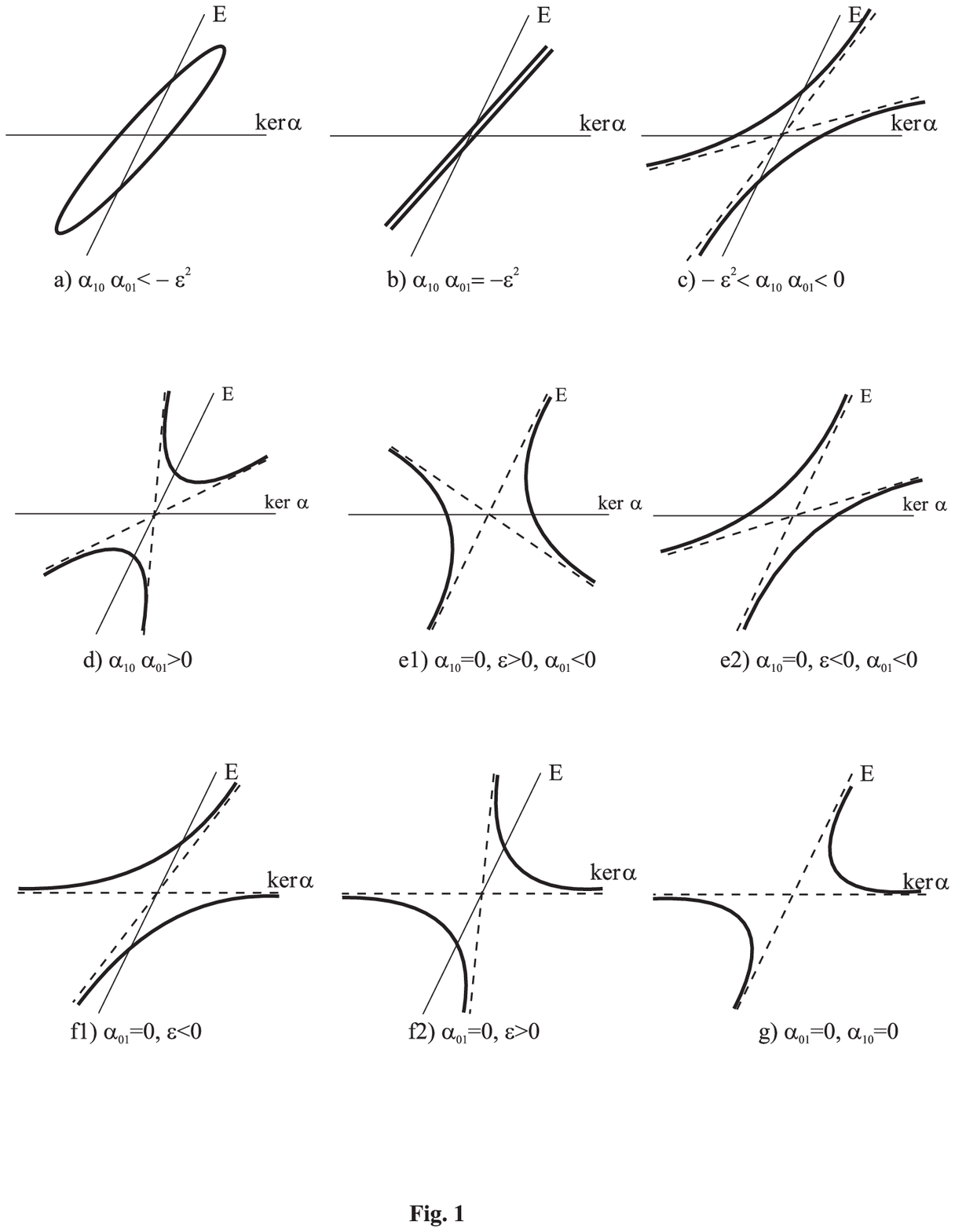}

\caption{Geometrical representation of possible special linear transformations
in $(1-1)$ space time}
\end{figure}
\end{document}